\documentclass{kapproc}
\usepackage[dvips]{graphicx}
\usepackage{procps}
\let\footnote\savefootnote

\let\footnotetext\savefootnotetext

\let\footnoterule\savefootnoterule

\let\lcitebracket[
\let\rcitebracket]
\begin{document}
\articletitle{Laser Control of Atomic Motion \\
inside Diatomic Molecules}
\chaptitlerunninghead{Laser Control of Atomic
Motion} \rhead{Laser Control of Atomic Motion}
\author{V.M.Akulin\altaffilmark{1}}
\altaffiltext{1}{Laboratory Aim$\acute e$ Cotton, Bat. 505,Campus d'Orsay,
91405 Orsay, France}
\author{V.A.Dubovitskii\altaffilmark{2}}
\altaffiltext{2}{Institute of Chemical Physics, 142432 Chernogolovka,
Russia}
\author{A. M.Dykhne \altaffilmark{3}}
\altaffiltext{3}{TRINITI, 142092 Troitsk, Russia}
\author{A. G.Rudavets\altaffilmark{4}}
\altaffiltext{4}{Institute of Chemical Physics, 142432 Chernogolovka,
Russia} \email{Arudavets@mics.msu.su\footnote{address for correspondences}}

\begin{abstract}
Globally optimal solution describing a phase conjugated field of Raman
scattering on the resonant  $B\gets X$ transition of iodine $I_2$ is
studied. Maximum optical coherence is found as a top eigenvalue problem.  A
reversibility theorem has been stated.  This provides sufficient conditions
for a tightly localized waveform and  molecular hologram to exist.  A noisy
picosecond pulse has been computed to show how femtosecond polarization is
regained  at target time.
\end{abstract}

\section{ Introduction}
Molecular wave packet engineering has attracted much attention in the works
collected under the rubric of "quantum control" $^{1-14}$. Current
world-wide efforts in the problematic have been mounted to develop efficient
methods for breaking selected molecular bonds  $^{11-14}$ or to harness
specific molecular states for optical processing devices and spectroscopic
uses. The rapid progress did not take a long time, because it was prepared
by the enormous lore in photomolecular spectroscopy accumulated since Lord
Rayleigh epoch and the beginning of quantum mechanics. From the other hand,
optimal control theory comprising dynamic programming and modern variational
calculus has been the subject of mathematical studies enabling to propose a
theoretical apparatus to the quantum control.

Of prime importance  were  the  minimum quantum uncertainty states
introduced by Schr\"{o}dinger.  For decades they were meant of only as
"Gedankenexperimente". Up-to-date femtosecond technique has made it possible
to observe both classically confined states of Rydberg atoms $^{4}$ and
space-localized wave packets in molecules $^{6,9}$.  The title of this paper
obliges us to restrict ourselves by the latter.  The quantum control theory
of ultrafast events close to dissociation limit of diatomic molecules is our
major concern. The challenge here is  to find optimal laser excitation of
picosecond scale causing femtosecond radiation  of an optically thin sample.

\medskip

We shall deal with the subject regarding the iodine $I_2$ molecule as our
test example.  Diatomic iodine in gas or condensed state has become the
reference standard $^{15}$ and ideal benchmark $^{16,17}$ in  modelling
the wave packet evolution.  Our task is greatly facilitated by the
considerable volume of researches, in which the quantun control of
molecular motion has been exploited theoretically $^{5,7,9}$. The
following experiments $^{6,8,9}$  with the laser-induced fluorescence
(LIF) have supported the idea of wave packet localization inside a
molecule. These works have again emphasized the link of the phase
modulated photoexcitation and molecular vibrations, which was broadly known
in Raman spectroscopy from 1920s $^{18}$.

\medskip

The goal yet achieved in the dynamic quantum control $^{9}$ includes
localization of the vibrational wave packet  at the attractive side of
molecular potential.  This scheme was referred to as the molecular
reflectron $^{5}$.  Bulk literature has been devoted to focusing the
matter states $^{3-9}$. The right posed theoretical limits $^{10,11}$
indicate an  existence of femtosecond coherence.

\medskip
Our approach to the problem of a drastical shortening optical transient
of a single molecule is started from a crucial relation between a
time-reverse molecular dynamics and phase conjugate resonance scattering
field.  A considerable interest presents a justification of the relation,
which, despite its generality, we have not been able to find in the
literature.  There are two aspects:  designing a right chosen
objective and  tailormaking  optimal optical pulses.  To understand
how they appear, it is worth noting, that the rapid improving of
femtosecond techniques is based on compression of chirped light pulses
provided by the well-established wave guides in optics. One can borrow the
key element of the pulse compression physics to apply it to squeezing an
optical coherence inside molecular space.  Namely, the Franck-Condon
region must play a role both dispersive wave guide and frequency modulator
owing to molecular vibrations.

\medskip

The vibrational wave packet is  expected to be focused at will on inner
Franck-Condon region, since  the Franck-Condon factor is at its maximum at
the steep repulsive curve.  Simultaneously the momentum variance $\Delta p
\approx 2Mv_0$ will be maximal at this point with  $v_0$ being the wave
packet velocity.  The uncertainty principle $\Delta p\Delta q \sim \hbar$
guarantees a tight localization $\Delta q $ of the rebound wave
packet.  Thus, its overlap with the ground state, having the variance
$\Delta q_0 \gg \Delta q$, lasts just for recoil lifetime ($\sim\Delta
q_0/v_0$) and gives rise to ultrashort coherent transient.  This picture
appeals to the "billard ball" model, $^{19}$ which still remains to be
extended to involve spreading wave packets on molecular curves. As example
we shall look at the reflectron scheme.

\medskip

The specific iodine reflectron $^{5,6,9}$ works at certain frequency above
the ground state X. A tailored electric field excites the B state until
the wave packet begins to concentrate near an outer turning point far
removed from the location of the original Franck-Condon transitions.  The
closer the excitations to the dissociation limit, the longer a delay
before the wave packet will be reflected from the outer curve and moves
back to inner repulsive core.  From here the wave packet recoils and may
radiate photons of resonant frequency to the B-X transition,  so that, an
optical coherence begins to appear with a some delay.  This "dark" period
might cause a fluorescence to spark at a chosen time.

\medskip

Having the name drawn from the electronic prototype, the molecular
reflectron is the scheme to create vibrational wave packet and to focus it
on a desired material target.  Our prime interest is a radiation process,
in which  the wave packet is periodically returning into resonance
Franck-Condon region and recoiling from it. Hence, a molecular "magnetron"
is the best-fitting term to our design, which underlines a parallel between
optical-molecular and radio-electronic phenomena.  In fact,  the
vibrational  wave packet spreading assisted by molecular anharmonicity
could be negated by the tailored laser pulse, because an interplay between
its modulation and quantum dispersion squeezes the wave packet, as  does a
properly chirped radio pulse propagating along a dispersive delay line.

\medskip

Molecular states promoted nearby their dissociation limit can be
localized more tightly  than the ground state by itself.  For the
excited levels are quasiclassical in their nature. Thus, we can treat
the recoil lifetime $\tau_{rec}$ as a kinematical overlap of the state X and
B presented  schematically in Figure 1. Here, the resonance Franck-Condon
transitions are drawn by the vertical arrows;  the horizontal arrows
designate the wave packet motion.  The wave packet velocity $v$ and recoil
time $\tau_{rec}$ are found from the potential functions (molecular
curves shown in Figure 1) as

$$
v(r)=\left[2(\varepsilon-U_{B}(r))/M\right ]^{1/2}, \qquad
\tau_{rec}\approx\int\limits^{R_0+\Delta R}_{R_0}\,dr/v(r).
$$
The distance $R_{0}=2.7 \AA$ is the equilibrium molecular position in the
ground X state of $I_2$. The  $\Delta R$ is a maximum variance for the
overlapping  states,  where $\varepsilon$ is the energy of excitation
(equal to an optical frequency) and $M$ being the reduced mass of $I_2$.
The variance $\Delta R$=$(\hbar/M\Omega)^{1/2}\approx 0.05\;\AA$ (with
$\Omega\approx 214 cm^{-1}$) for the ground state is more than it for
focused B wave packet on the inner molecular wall.  Furthermore, if its
variance is disregarded, the recoil lifetime is limited below by the
magnitude $\tau_{rec} \approx 15\; fs$.  The state B  can be populated at
the energy $\varepsilon\approx 19050 cm^{-1}\; (\sim 525\; nm) $ for a
more long time than the recoil time $\tau_{rec}$.  Setting the gate of
excitation by the typical vibrational period $T \approx546\; fs$ at
$\varepsilon$, one can measure the squeezing by the ratio $T/\tau_{rec}$
which is expected about 30 or more. As it will be clear later, even subtle
details of controlling the atomic motion inside diatomic molecules can be
understood by means of the classical characteristics and semiclassical
distribution functions.

\medskip

The plan of our paper is as follows. In section 2 we apply the optimal
control theory to optical polarization in the weak field response. "A
priori"  pulse shape will not be conjectured.  An exact and unique
solution to a maximum coherence peaked at target time is derived for
isolated $I_2$ molecule in optically thin media. A time reverse (phase
cojugate) resonance Raman scattering providing for a feedback to the
vibrational spreading is represented as a basic principle of the quantum
control.  Controlling the pure and mixed states is considered.  In section
3 we reconcile the quantum machinery with a reasoning appealing to the
classical kinematics  and discuss the phase modulated field of the
scattered radiation associated with the classical action of the recoiling
wave packet. Section 4 illustrates the general formulation by the
numerical simulation. In section 5 we conclude and outline the future
prospects.

\medskip

\section{ Mapping the Matter to a Light Field}

A rigorous  mathematical treatment of molecular "magnetron" necessitates
to solve the  quantum equations dealing  on equal footing  with vibrational
motions and nonadiabatic electronic transitions.

\medskip

The adiabatic dynamics  of a diatomic molecule   is governed by
Hamiltonian operators $\hat H_{b}=\hat T_{kin}+\hat V_{b}$ and $\hat
H_{x}=\hat T_{kin}+\hat V_{x}$.  Here $\hat T_{kin}$ is the operator of
radial kinetic energy, where the only active coordinate is a distance
between atoms,  $\hat V_{b}$ and $\hat V_{x}$ model the adiabatic
potentials of the B  and X states respectively. Having the fastest
vibrational motion in diatomic molecule, we neglect the  more slow
molecular rotations (and the fine structure as well), since those
transients are separated by time scales $^{15}$.  For heavy molecule as
$I_2$, the vibrational transient occurs on a subpicosecond time, while
the rotational transient occurs on a longer time scale of 10 ps.  Then, we can
limit our consideration by the former, because the angular correlations
happen past the vibrational ones. In fact, the rotational correction will
only require a modification of the Hamiltonians and averaging over initial
rovibrational states.  As justified in the works $^{7,9}$, the inclusion
of the molecular rotations does not abandon the dynamic quantum control of
vibrational wave packet towards a desired goal.

\medskip

Throughout this paper the Condon approximation for dipole transition
moment $\mu$ is used, with $\mu$ being independent of internuclear
separation in molecule. This assumption is valid for the weak field
quantum control, when the molecular ground state is being well localized.
The electronic states are coupled by the electric dipole operator $\hat
V_{bx}(t)={\hat \mu } {\cal E}(t)$, where the laser field is
$$
{\cal E}(t)=E(t)e^{{i}\omega_{0}t/\hbar}
+\stackrel{*}{E}\!(t)e^{-{i}\omega_{0}t/\hbar}
$$
and its slow envelop $E(t)$ of allocated duration $T_{p}$ is to
be found under a constrain limiting the pulse energy,
\begin{equation}
J=\int\limits_{0}^{T_{p}}d\tau\;E(\tau)\stackrel{*}{E}\!(\tau).
\end{equation}
This laser field should prepare the molecular transition to
a desired goal. Let optical dipole target be a linear
superposition of weighted Dirac $\delta$-like envelopes shifted on
$t_{\alpha} $,
\begin{equation}
G(t)=\sum_{\alpha}G_{\alpha}\delta(t-t_{\alpha}).
\end{equation}
The wisdom of that representation to the resonance coherence
will be obvious further.

\medskip

The Schr\"{o}dinger equation for the wave functions $\Psi_{b}$
and $\Psi_{x}$ reads as
\begin{eqnarray}
i\hbar\dot\Psi_{b} & =&(\hat H_{b}-\omega_{0})\Psi_{b} + \hat V_{bx}(t)\Psi_{x}
\nonumber \\
i\hbar\dot\Psi_{x} & = &\hat H_{x}\Psi_{x}+\hat V_{bx}^{\dag}(t)\Psi_{b} .
\end{eqnarray}
The empty B state and populated ground X state (at zero temperature) will
be of use as our initial condition $\Psi_{b}(0)=0, \quad
\Psi_{x}(0)=\Psi_{x,0}$.  Also, the rotating wave approximation is
utilized in eq (3) to avoid fast oscillations of optical
frequency of the B-X transition. Then, the resonance interaction is given
by the slow amplitude $ \hat V_{bx}(t)= {\hat \mu } \stackrel{*}{E}(t)$.

\medskip

Our objective consists in controlling the polarization,
$$ D_{xb}(t)=\mu({\Psi}_{x}^{\dag}\vert\Psi_{b})=\mu\int
\limits_{0}^{\infty}\,dr\stackrel{*}{\Psi}_{x}(r,t)\Psi_{b}(r,t),
$$
which in its turn manifests in the molecular optical susceptibility.  The
first correction to the transition dipole moment $D_{xb}(t)$ in the weak
field regime is
\begin{equation}
D_{xb}(t)=-i\int\limits_{0}^{T_{p}}d\tau {\cal S}_{00}(t-\tau)
\stackrel{*}{E}(\tau) , \;\;\;\;\;\;\;\;\;\; (t>T_p).
\end{equation}
To characterize the molecular transition, it is useful to extract a
temporal structure factor independing of the excitation field envelope
$E(\tau)$. For this goal we introduce a wave-packet correlation function as
\begin{equation}
{\cal S}_{00}(\tau)=\frac{\mu^{2}}{\hbar} (\Psi_{x,0}^{\dag}
\vert e^{-{i}\tau\hat H_{b}/\hbar}\vert\Psi_{x,0})
\end{equation}
This formula describes the optical polarization caused by the $\delta$-like
pulse of electric field $E_{s}(\tau)=\delta(\tau)$.  Modulation of the B-X
transitions is formed while the $\Psi_{x,0}$ replica propagates on the B
curve.  The bound wave packet oscillates spreading between turning points
and permits the molecule to radiate the Raman scattering signal.  In
frequency domain, the erratic spectral pattern $^{17}$ of the resonance
Raman intensity is accordingly    observed. The spectral profile of the
${\cal S}$ correlator is represented by the KHD formula for the fundamental
Raman overtone. Its resonance dependence is given by

\begin{equation}
I_r(\omega)\propto\Bigl|\int\limits_{0}^{\infty} d\tau e^{{i}\omega
\tau/\hbar}{\cal S}_{00}(\tau)\Bigr|^2 =\frac{\mu^{4}}{\hbar^{2}}
\Bigl|\sum\limits_{m}\frac{(\Psi_{x,0}^{\dag}\vert\Psi_{b,m})(\Psi_{b,m}
^{\dag}\vert\Psi_{x,0})}{\gamma+{i}(\beta_{m}-\omega)}\Bigr|^2,
\end{equation}
where $\gamma$ is the damping constant; the eigenstates
$\Psi_{b,m}$ and eigenleveles  $\beta_{m}$ can be found from
eigenequation $\hat H_{b}\Psi_{b,m}=\beta_{m}\Psi_{b,m}$.
The Stokes overtones are manifested when the wave packet reaches a
favorable position to overlap the vibrational X states. $^{21}$

\medskip

To maximize the overlap between the Franck-Condon density
$\vert D_{xb}(t)\vert^2$ and optical dipole target $G(t)$,
we define the field functional as
\begin{equation}
F=\int dt G(t)\vert D_{xb}(t) \vert^2= \int\limits_{0}^{T_{p}}d\tau
\int\limits_{0}^{T_{p}}d\tau_1 {\cal P}_{00}(\tau,\tau_1)
\;E(\tau)\stackrel{*}{E}\!\!(\tau_1).
\end{equation}
One should note, that given functional does not confine the optical polarization
throughout the time. However, it does guarantee a maximal dipole moment at
our objective $G(t)$. The Frank-Condon transitions develop freely beyond
the target time constrained only by the field energy in eq 1.  The
dynamic quantum control begins with  the X state $\Psi_{x,0}$, which is
tightly localized in the I-I distance at the start.  The B state must be
optimally driven below the dissociation limit to avoid bound-free
transitions resulting in losses.  Evident wave nature will impede the
control, in which the B wave packet having undamped oscillation  between
turning points,  must be focused on the inner steep core with a maximum
velocity at target time.  Those requirements will be met under a global
maximum of the field functional $F/J=\lambda$. This condition is set by
small variation of the probe field $\stackrel{*}{E}(\tau)$ in the
variational equation,
\begin{equation}
\delta(F-\lambda J)=0,
\end{equation}
where $\lambda$ is the Langrange multiplier enabling to enforce
the energy constrain.  Herefrom the basic problem of dynamic quantum
control in the weak field regime  reduces to the Fredholm
eigenequation,
\begin{equation}
\int\limits_{0}^{T_{p}}d\tau_1 {\cal P}_{00}(\tau,\tau_1)
E(\tau_1)=\lambda E(\tau),
\end{equation}
where the kernel of the homogeneous equation is given by the
complex hermitian matrix,
\begin{equation}
{\cal P}_{00}(\tau,\tau_1)= \int dt G(t) {\cal S}_{00}
(t-\tau){\stackrel{*}{\cal S}}_{00}(t-\tau_1).
\end{equation}
The integral kernel ${\cal P}$ depends exclusively on a chosen dipole
target and specification of the electronic transition being a
pump-independent in the weak field limit. In  some sense, it copies the
material properties of the molecule to a light field.  The time-dependent
matrix elements eq (5) forming the kernel in eq (9) have long been known
in the "matter-radiation" interaction theory, benefited to understanding
nonlinear
optics and utilized in quantum control of wave packets with different
material objectives: the minimum space variance of wave packets
(I.S.Averbukh, M.Shapiro $^{3}$), the delta-like space density target
(V.Dubov, H.Rabitz $^{ 7}$), the minimum quantum uncertainty state
(K.R.Wilson et al. $^{5,6,9}$).

\medskip

To recall their argumentation, we shall take a quick look at another
field functional, which describes a total B population in the weak field
regime,
\begin{equation}
N=\int\limits_{0}^{\infty}\,dr \vert \Psi_{b}^{(1)}(r,t)\vert ^{2}=
\int\limits_{0}^{T_{p}}\int\limits_{0}^{T_{p}}d\tau d\tau_1 {\cal M}_{00}
(\tau,\tau_1)\;E(\tau_1)\stackrel{*}{E}\!\!(\tau),
\end{equation}
where the self-conjugate  kernel  ${\cal M}_{00}(\tau,\tau_1)$ in
eq 5  is proportional to the ${\cal S}$-correlator depending on the
difference argument  $\tau-\tau_1$,  because the molecular potentials are
independent of time
\begin{equation}
{\cal M}_{00}(\tau,\tau_1)= {\cal S}_{00}(\tau_1-\tau)/\hbar
=\stackrel{*}{\cal S}_{00}(\tau-\tau_1)/\hbar
\end{equation}
Again one can repeat above steps searching a global maximum for
the functional $N/J=\lambda$, which gives a maximum population
yield per unit field energy. Considering a small field variation
$\stackrel{*}{E}(\tau)$ for the variational equation
$\delta(N-\lambda J)=0 $, we obtain an eigenequation
\begin{equation}
\int\limits_{0}^{T_{p}}d\tau_1 \stackrel{*}{\cal S}_{00}(\tau-\tau_1)
E(\tau_1)=\hbar \lambda E(\tau)
\end{equation}
We obtain the degenerated kernel having  separated
the time arguments $\tau$ and $\tau_1$
by means of the unity decomposition
$ \sum\limits_{m}\Psi_{\alpha,m}\Psi_{\alpha,m}^{\dag}=1$.
Then, a solution to the integral equation can be cast as

\begin{equation}
E(\tau)=\frac{\mu^2}{\lambda_{max}\hbar^{2}} \sum\limits_{m}
(\Psi_{x,0}^{\dag}\vert\Psi_{b,m})\,e^{{i}\beta_{m}\tau /\hbar}
{\cal E}_{m}
\end{equation}
Unkno wn coefficients  ${\cal E}_{m}$  are found from the
system of linear equations taken at a maximal eigenvalue
$\lambda_{max}$
\begin{equation}
{\cal M}_{m,n}{\cal E}_{n}=\lambda_{max}{\cal E}_{m}
\end{equation}
The matrix  ${\cal M}_{m,n}$ is formed by the Franck-Condon
factors and energy levels of the B state
\begin{equation}
{\cal M}_{m,n}=(\Psi_{b,m}^{\dag}\vert\Psi_{x,0})(\Psi_{x,0}
^{\dag}\vert\Psi_{b,n})\frac{e^{{i}(\beta_{m}-\beta_{n})
T_{p}/\hbar}-1}{{i}(\beta_{m}-\beta_{n})}
\end{equation}

\medskip

The optimal field $E(\tau)$ depends on the oscillator strengths of the
molecular transition in eq(16), which includes factors of the increasing
frequency due to the energy ladder $\beta_{0} <\beta_{1}...<\beta_{m}<...$
of vibrational levels.  This field  must steadily populate the molecular
states.  It is expedient to put on the textbook Condon's model of the
molecular  transition between the parabolic curve $V_x=M\Omega^2 r^2/2$
and flat continuum $V_b=const$, where the S-correlator  is
the following:

$$
{\cal S}_{00}(t)=(1+i\Omega t/2)^{-1/2}=A_0(t)\;e^{-i\phi_0(t)/2}
$$
The phase profile $\phi_0(t)=arctg(\Omega t/2)$ is provided by the
spreading wave packet in continuum which is projected to the ground
vibrational state. On using a slope molecular curve $V_b=-f r$ models
the repulsive Franck-Condon region and  recoiling wave packet.
The phase correction $\phi(t)=\phi_0(t)+t(ft)^2/(12M\hbar)$ is explicitly
derived in the model dealing with the momentum representation of wave packet dynamics. The amplitude correction turns out to be of Gaussian type and like the  Debye - Waller factor is  given by the ratio of the recoil energy $R=(ft)^2/(2 M)$ to
vibrational quantum $\hbar\Omega$, i.e.

$$
A(t)=A_0(t)e^{-R/(2\hbar\Omega)}.
$$
The flat and slope continuum of molecular states provide for the  kernel
${\stackrel{*}{\cal S}}_{00}(\tau-\tau_1)$ with negative chirp of frequency
for a short time duration $\tau_1$, when the expansion in the power
of $\tau_1$ is legitimate.  The conclusion holds also in a general
picture of semiclassical approximation as demonstrated in next section.
Thus, a short pulse having positive  frequency chirp slows down the fast
integrand oscillations in eq (13) enabling one to maximize its eigenvalue,
(i.e. the yield of the population). The maximum population has to involve a
gain of fluorescence.$^{22-24}$

\medskip

Other trends of wave packet correlations make it possible to control a
selected molecular state or coherent transients.  According to eq 9, a
top eigenvalue $\lambda$ gives a maximum optical dipole transition per
unit of field energy having eigenvector $E_{\lambda}(\tau)$ as the
globally optimal field of the allocated duration $T_{p}$. To demonstrate
this statement we address the $\delta
$-Dirac model as the simplest optical dipole target,
where the coherent envelope $G(t)=\delta(t-t_{d})$ will be peaked
at time $t_{d}$ after turning off the excitation pulse.  The condition
$t_{d}>T_{p}$ means, that the "spontaneous" polarization will stand out on
the pulse-free background.  The delta-like target model results in the
degenerate Fredholm kernel.  An unique solution to the integral equation
is explicitly written as

\begin{equation}
E_{\lambda}(\tau)={\cal S}_{00}(t_d-\tau)
(J/\lambda)^{1/2},
\end{equation}
where $t_{d}$ marks a moment when the optical polarization is regained with
$\lambda$ being the normalized coherent yield,
\begin{equation}
\lambda=\int\limits_{0}^{T_{p}}d\tau\stackrel{*}
{\cal S}_{00}(t_d-\tau) {\cal S}_{00}(t_d-\tau).
\end{equation}
The indication of optimality according to Bellman principle is indeed
realized: the control depends on the state of system at the current moment
alone. This globally optimal fields can be understood in general terms,
for their envelopes $E_{\lambda}(\tau)$ match those of the wave packet
correlations representing the time-reversed resonance scattering.  The delay
time $t_d$ fits an absolute maximum of the optical transient to target
time $\tau=t_d$. The dipole transition moment is given by the convolution
integral

\begin{equation}
D_{xb}(t)=-i
\left(J/\lambda \right)^{1/2}\int\limits_{0}^{T_{p}}d\tau
{\cal S}_{00}(t-\tau)\stackrel{*}{\cal S}_{00}(t_d-\tau).
\end{equation}
The optimal field in eqs 17, 19 cancels out the fast oscillating
behaviour of the integrand as being designed to have the phase conjugate
wave-packet correlations excited by the $\delta$-like pulse.
The optical transient by virtue of eq 19 exhibits periodical recurrences,
having the maximum magnitude $-i(J\lambda)^{1/2}$ exactly at the target time
$t_d$.

\medskip

The straightforward extension of the $\delta$-dipole target to the
realistic shape $G(t)$ specifies an eigenvalue problem in eq 9.  The
integral operator ${\cal P}$ can be decomposed into sum of the degenerate
kernels according to eq 10 in the weak field limit.  The solution is
\begin{equation}
E(\tau)=\sum_{\alpha}{\cal S}_{00}(t_{\alpha}-\tau)
G_{\alpha}{\cal E}_{\alpha}/\lambda_{max}
\end{equation}
where the ${\cal E}_{\alpha}$ being the eigenstate at a top eigenvalue of
the system
\begin{equation}
{\hat {\cal P}}_{\alpha \alpha_1} {\cal E}_{\alpha_1}=\lambda_{max}
{\cal E}_{\alpha} \,
\end{equation}
The  matrix coefficients  are given by,
\begin{equation}
{\hat {\cal P}}_{\alpha \alpha_1}=G_{\alpha}
\int \limits_{0}^{T_{p}}d\tau {\cal S}_{00}(t_{\alpha}-\tau)
\stackrel{*}{\cal S}_{00}(t_{\alpha_1}-\tau)\,
\end{equation}
so that the optimal field satisfies the energy constrain by
the definition.

\medskip

We have established an important fact, which merits to be reformulated as
a reversibility  theorem  asserting  sufficient conditions to reverse the
spread wave packet in time:
{\it In order to drive molecular transition to the polarization target
$G(t)$, the optimal field  should be made of superimposed phase conjugate
fields of the resonance Raman scattering of ultrashort pulses of
amplitudes $G_{\alpha}{\cal E}_{\alpha}/\lambda_{max}$ at the delay time
$t_{\alpha}$.}

\medskip
The resonance scattering radiation itself may seem to be  "erratic" in
time.  The wave-packet collapse and spreading  and destructive interference
are responsible for  the apparent noise  due to molecular "disorder" on
account of potential anharmonicity, uncommensurate frequencies, curve
crossing etc.  But phase conjugate feedback with respect to molecular
correlations (resonant scattering) allows the wave packet to be recovered
at a fixed space-time point at will.  The optical coherence can be
restored not only for the Raman fundamental overtone. Our treatment might
be readly extended  to controlling the high overtones of molecular
resonance scattering.

\medskip

We term the field enabling to squeeze the optical coherence as a quantum
hologram to note its key role for a wave packet interference
in the Franck-Condon region.  The quantum holography
is relying on the wave packet correlations, which can be found by
detecting a fluorescence or population created by two phase locked
ultrashort pulses.  Accordingly, in the limit of weak field,
the excited state B is obtained from eq  3, in the first order to
the resonance interaction as,
\begin{equation}
\hat\Psi_{b}^{(1)}(t)=-{i}\theta_{1} e^{-{i} t \hat H_{b}/\hbar}\hat\Psi_{x,0}(0)
-{i}\theta_{2} e^{-{i} (t-T) \hat H_{b}/\hbar}\hat\Psi_{x,0}(T).
\end{equation}
This superposition is obvious analog of an incident and object waves in
optical holography. To produce the quantum interference of wave packets,
the first laser pulse at $t=0$ must be followed by  the  delayed pulse at
$t=T$. If the laser pulses are phase-locked and their areas are small with
respect to $\pi$, the population $N$ of the state B contains the
contribution $^{14}$  of the one photon two-pulse interference, i.e.
\begin{equation}
N=\int\limits_{0}^{\infty}\,dr \vert \Psi_{b}^{(1)}(r,t)\vert ^{2}
=\vert\theta_{1}\vert^2+\vert\theta_{2}\vert^2+
Re\{\,\theta_{2}\stackrel{*}\theta_{1}{\cal S}_{00}(T)\,\}
\end{equation}
The alternative pathways in course of the Franck-Condon transitions depend
on the delay time $T$ between pulses.  This dependence is being just
required to controlling the wave packet motion.  The phase
locked laser pulses  must retrieve synphase and quadrature components of
the $\cal S$ correlator. The LIF signal $^{14}$, ionization channel
$^{25}$, or other means for the matter-wave interferometry may be employed.
Then, the use of programmable optics $^{26,27}$ and algorithms based on
the reversibility theorem fed into computer codes can be made to tailor
the optimal field.  This quantum holography program must be realized to
rebuild the well localized replica of the ground state or to squeeze
optical coherence as called on even for unknown molecular curves as well as
collisional or intramolecular dephasing and relaxation.

\medskip

The reversibility theorem still holds for mixed states inevitably residing
in statistical systems, for which irreversible processes hamper the
quantum control of the wave packets.  By considering the nondiagonal
density matrix of the resonance transition $\hat\rho_{\alpha
\beta}=\langle\hat\Psi_{\alpha}^ {\dag}\hat\Psi_{\beta}\rangle$, the
brackets $\langle...\rangle$ symbolize an ensemble averaging in
quantum kinetic theory.  Here, the populations
$\hat\rho_{xx}=\langle\hat\Psi_{x}^{\dag}\hat\Psi_{x}\rangle$,
$\hat\rho_{bb}=\langle\hat\Psi_{b}^{\dag}\hat\Psi_{b}\rangle$ and
polarization $\hat\rho_{bx}=\langle\hat\Psi_{b}^{\dag}\hat\Psi_{x}\rangle$
adhere to the kinetic equations:
\begin{equation}
 i\hbar\frac{\partial\hat \rho}{\partial t}=
\left[{\hat{\cal H}},{\hat \rho}\right]-
i\left(\Gamma {\hat \rho} \right)
\end{equation}
where the rectangular brackets $[...]$ denote a quantum commutator
$\hat{\cal H}{\hat \rho}-{\hat \rho} \hat {\cal H}$ of the operators
which are represented in the matrix notation as:
\begin{equation}
 {\hat \rho}=\left( \begin{array}{c}{\hat\rho}_{bb},\;\;{\hat\rho}_{bx} \\
  {\hat \rho}_{xb},\;\;{\hat\rho}_{xx}\end{array}\right),\;\;\;\;\;\;
  {\hat {\cal H}}=\left( \begin{array}{c}{\hat H}_{bb},\;\;{\hat V}_{bx} \\
  {\hat V}_{xb},\;\;{\hat H}_{xx}\end{array}\right)\;\;\;\;\;\;\;\;\;\;\;\;\;\;
\end{equation}
A phenomelogical damping matrix $\hat \Gamma$ designates the overall
losses in irreversible processes for collisions, spontaneous radiation,
etc. Experiments in solids, liquids and gas cell conditions indicate
an existences of coherent vibrational transients of $I_2$ to tens of
picoseconds $^{14}$.  Thus, the quantum control theory within the matrix
density formalism, in which the relaxation will not be scrutinized
further, may be applied.

\medskip

Just as for pure states, we take care of the optical polarization
${\hat\rho}_{bx}$ at target time. One can immediately maximize  the
overlap between the optical transition ${\rho}_{bx}$ and its dipole
target in eq 2. By letting the polarization  operator ${\hat \rho}_{bx}$
be  a complex valued quantity, we define an overlap functional  as
\begin{equation}F=-\mu\int dt\; Im\{G(t)\;Tr\{\hat\rho_{bx}\}\}=\int_{0}^{T_p}
d\tau\; Re\,\{\stackrel{*}{E}(\tau) {\cal P}_{00}(\tau)\}
\end{equation}
where another field functional ${\cal P}_{00}(\tau)\}$ is easily obtained
from Eqs. (25),(26) to be
\begin{equation}
{\cal P}_{00}(\tau)=\frac{\mu^{2}}{\hbar}\int dt\; G(t)
e^{-\Gamma (t-\tau)}\; Tr\;\{ e^{-i(t-\tau){\hat H}_{b}/\hbar}\,
[{\hat \rho}_{xx}-{\hat \rho}_{bb}]\, e^{i(t-\tau){\hat H}_{x}/\hbar} \}
\end{equation}
The functional $F$ is indentified as a work that a given target $G(t)$
would produce on the Franck-Condon  transition. This statement can be
readily understood, since according to eq 25 the field functional $F$
can be rewritten  as the populations difference due to the target field
"action"
$$
F=-\int dt\; Im Tr\left [ G(t){\hat \rho}_{bx}(t) \right ]=
\hbar \int dt\; Tr\left [{\dot{\hat\rho}}_{b}(t)-{\dot{\hat\rho}}_{x}(t)\right ]
=\hbar (N_b-N_x)
$$
To maximize F, we apply the standard variational procedure of eq 8.  The
dynamic quantum control in the weak field limit   reduces to a problem of
a linear mathematical programming. As a bonus of the approach, no integral
equation needs to be solved at all, because, even for any target shape
$G(t)$, the variational equation
$$
\delta(F-\lambda J)/\delta\stackrel{*}{E}(\tau)=0
$$
results in an unique solution:
\begin{equation}
E(\tau)={\lambda^{-1}}{\cal P}_{00}(\tau)
\end{equation}
By using the empty B state and populated state X, ${\hat\rho}_{bb}=0\;\;
{\hat \rho}_{xx}={\hat\rho}_{0},\;\; $  as the initial conditions, we set
that
\begin{equation}
{\cal P}_{00}(\tau)=\frac{\mu^{2}}{\hbar}\int dt\;
G(t) e^{-\Gamma (t-\tau)} Tr\,\{ e^{-i(t-\tau){\hat H}_{b}/\hbar}
{\hat\rho}_{0} e^{i(t-\tau){\hat H}_{x}}\}
\end{equation}
and the maximum coherent yield $\lambda$ becomes explicitly  known from the
energy constrain represented as
\begin{equation}
\lambda^2=
{J^{-1}\, \int_{0}^{T_p}\stackrel{*}
{\cal P}_{00}(\tau){\cal P}_{00}(\tau)} d\tau.
\end{equation}
By choosing the pure state ${\hat\rho}_{0}=\Psi_{x,0}^{\dag}\Psi_{x,0}$
for the initial condition, we recover the  main result of eq 5, as it
must.  Matrix density formulation has the advantage of being applied to
irreversible physical systems of gas and condensed matter.  The
predissociation \footnote{ J.Jortner has drawn our attention to curve
crossing effects in experiments reported by V.Apkarian [8] et al. and
M.Chergui [17] et al. Our theory can be extended for B-(B",a,a')
couplings affecting the quantum control. The results is the entanglement
with continuum molecular states shortening optical transients.} and caging
phenomena can be
considered as well. What is more, the observation and control of wave
packets is in the progress $^{8,9}$ and includes rearrangement of subtle
chemical structures in solids and liquids.  Having postponed the feasible
generalization, we shall discuss in next section the reversibility theorem
in the semiclassical uniform approach.

\section{ Semiclassical approximation}

The semiclassical limit of quantum dynamics is characterized by
classical equations, that require to be supplemented by the quantum
initial conditions to take full account of the singularity $\hbar=0$.  The
former is clear, only classically possible paths joining $R_{0}$ with
$R_{1}$  contribute to the ${\cal S}$-correlator. These paths are invoked
by Hamilton's principle $\delta{\cal W}=0$, where the classical action is
\begin{equation}
{\cal W}(R_{0},R_{1},t)=
\int_{0}^{t}d\tau\left(\frac{M}{2} \dot R^2(\tau)-U_b(R(\tau))\right)
\end{equation}
If the action is known, it enables us to find a momentum $P_0=-\partial
{\cal W}(R_{0},R_{1},t)/\partial{R_0}\quad$ and density  of states $\partial
{P_0}/\partial {R_1}$.  Appearance of the derivative of classical
trajectory in determining the density distribution in phase space
is not occasional $^{28}$.
\medskip
The quantum transition amplitude connecting the points $R_{0}$ and $R_{1}$
depends on Plank 's constant $\hbar$ and can not be deduced from the
classical postulate. Instead, the quantum grounds dictate
\begin{equation}
{\cal K}(R_{0},R_{1},t)=\langle R_{0}\vert
e^{-i t H_{b}/\hbar}\vert R_{1}\rangle=
(-2i \pi\hbar \delta {R_1}/{\delta {P_0}} )^{-1/2}
\exp (i{\cal W}(R_{0},R_{1},t)/\hbar).
\end{equation}
The probability density, $\vert{\cal K} \vert^{2}=(2\pi\hbar)^{-1} {\delta
P_0}/{\delta R_1}$, stems from the total number of atoms reached the point
$\delta R_1$ started out from the point $R_0$  and having the classical action
${\cal W}={\cal W}(R_0,R_1,t)$. The eq(33) does not give the quantum
amplitudes correctly for certain space-time points.  Suffering from known
drawbacks, it fails in the classically forbidden region, caustics and
turning points. The  story is old as the quantum theory itself.  However,
for our goals, the semiclassical approach does at least serve to  a clue
idea about the quantum control  $^{29}$.

\medskip

The classical action  $W$  is inherited by the phase of electric field
promoting resonant B-X transition.  By this, the optimal laser field is
related with the S-correlator which projects the wave packet B to the
ground state X.  The  latter is centered at $R_0=2.7 \AA$ in the I-I
distances  with the variance $\Delta R=0.05 \AA$ being much less than the
Franck-Condon region extension itself.  Hence, the mere closed paths
passing the point $R_0$ contribute to the optimal field. We can present
its phase by the classical action eq 33 expanded for ultrashort pulse
as

\begin{equation}
{\cal W}(R_{0},R_{0},t_d-\tau)={\cal W}(R_{0},R_{0},t_d)+U_{b}(R_{0})
\tau-M(\ddot R_{0})^{2}\tau^{3}/3.
\end{equation}
in the vicinity of inner turning point $R_0$.  The potential $U_{b}(R_0) $
plays the role of frequency detuning of the B-X
transition. Owing to zero velocity $\dot R_0=0$,  its linear chirp will be
negligible as compared with the nonlinear chirping. Instead,
there is a significant quadratic chirp $-M(\ddot R_{0})^{2}\tau^{2}$t
as adominant feature of  acceleration $\ddot R=-M^{-1}(d U_{b}(R)/d{R})$
due to the steep inner core. By travelling a distance $\delta r$ in vicinity
of $R_0 $, the wave packet detunes down the resonant frequency on $ \delta
r (d U_{b}(R)/d{R})$. The greater $\delta r=\ddot R_{0} \tau^2$ the
smaller the frequency of B-X transition.  This gives the negative chirp,
which qualitatively is in agreement with the known reasoning $^{6,29}$. In
practice, for a more longer pulse duration, this simple classical expression
may overestimate the chirp rate,   because the particle also travels far
from the steep potential in a less repulsive Franck-Condon region.
Reducing of frequency modulation, as well as linear chirping owing to a
wave packet velocity is quite possible.

\medskip

A care must be taken to the turning points, where the semiclassical
transition amplitude postulated in eq 33 diverges ${\partial
R_1}/{\partial P_0}=0$. This is the case for our molecular magnetron
design, in which  the wave packet must be focused just in vicinity of the inner
turning point. This severity can be circumvented by the uniform approach
based on the Wigner representation of quantum operators. To find the
S-correlator density, we must overlap the initial and propagated
distributions in the phase space $s=(r,p)$,
\begin{equation}
\vert{\cal S}(-t) \vert^2 =\frac{\mu^4}{\hbar^2}
\int\limits_{s} {drdp}\;\rho_{0}(r,p)\;\rho_{0}(R,P)
\end{equation}
where $ R=R(-t;s,0)$ and $P=P(-t;s,0)$ being the current coordinate and
momentum of a particle experiencing the force $-\partial U_b(R)/\partial
R$. The classical integrals of motion are fixed by the initial conditions
$R(0;s,0)=r$ and $P(0;s,0)=p$ in the phase space. According to the Newton
laws, the particle travels as,
\begin{equation}
{\dot R} = {P}/{M}\quad {\dot P} = -\partial U_b(R)/\partial R,
\end{equation}
and the Wigner density function obeys the transport equation:
\begin{equation}
\frac{\partial \rho_{0}(s,t)}{\partial t}+\frac{p}{M}
\frac{\partial\rho_{0}(s,t)}
{\partial r}-\frac{\partial U_b(r)}{\partial r}\frac{\partial
\rho_{0}(s,t)}{\partial p}=0.
\end{equation}
Hence, the solution  is $ \rho_{0}(R(0;s,t),P(0;s,t))$ propagating
along the trajectory passing the point $s$ in time $t$ with the distribution
for the ground X state, which is defined as
\begin{equation}
\rho_{0}(r,p)=\int\limits_{-\infty}^{\infty}\frac{dq}{2\pi \hbar}
e^{i qp/\hbar}\Psi^{\dag}_{x,0}(r+\frac{q}{2})
\Psi_{x,0}(r-\frac{q}{2})=\frac{1}{\pi \hbar} e^{-F(p,r)},
\end{equation}
where
\begin{equation}
F(p,r)= ((r-R_0)/\Delta R)^2+(p \Delta R/\hbar)^2.
\end{equation}
The Wigner function of the ground state X is sharply  peaked at the point
$q_0=(R_0,0)$ having Gaussian variances $(\Delta R,\hbar/\Delta R)$ in
the coordinates and momenta, respectively.  The classical orbits passing this
stretched ellipse make a major contribution to the ${\cal S}$-correlator
amplitude.  \footnote{It is worth noting, that a central role of the classical
trajectories  for calculating matrix elements of high excited states has
been well known since the thirties, due to Landau seminal contribution to
the chemical collisions theory $^{30}$.}  Then, one can apply the fastest
descent method to estimate  the amplitude in eq 35. We expand
the exponent functional $F(R,P)$ to second order in the declines
from the path $s_0(t)=(R(0;q_0,t),P(0;q_0,t))$ passing
the equilibrium point  $q_0$  of the ground X state.  The family of
orbits in its neighborhood contributes significantly. Thus, this topology
determines basically a quadratic form
\begin{equation}
F(R,P)=F^0+F^{0}_{s}\;(s-q_0)+\frac{1}{2}F^{0}_{s,s_1}(s-q_0)(s_1-q_0)+...,
\end{equation}
The function $F^0=F(s_0(t))$ is taken on the path $s_0(t)$. A small
stirring of its initial condition yields the four first and six second
derivatives of $R,\; P$ with respect  to  two-dimensional indices
$s=(r,p)$ or $s1=(r,p)$. Then, the quadratic form is settled by the twelve
related functions: $$ R(0;q_0,t), P(0;q_0,t), R_{s}(0;q_0,t),
P_{s}(0;q_0,t), R_{s,s_1}(0;q_0,t), P_{s,s_1}(0;q_0,t), $$ Of course, these
functions adhere to Newton's equations and their derivatives.
The useful vector and matrix notations
\begin{equation}
{\vec V}=(F^{\;0}_{r}, F^{\;0}_{p});\;\; {\hat {\cal F}}=
\left(\begin{array}{cc}(\Delta R)^{-2}+F^{\;0}_{rr}/2\;,&
\qquad\;F^{\;0}_{rp}\\
F^{\;0}_{rp}\;,&(\Delta R/\hbar)^{\;2}+F^{\;0}_{pp}/2
\end{array}\right)\;\; ,
\end{equation}
simplify the resulting Gaussian integral:
\begin{equation}
{\vert{\cal S}(t) \vert^2}\propto\hbar^{-1} (\det{\cal F})^{-\frac{1}{2}}
\exp\left(-F^0+\frac{1}{4}\vec V
{\hat{\cal F}}^{-1}\vec V^{\dag} \right).
\end{equation}
The derivatives of the classical path with respect to its starting point
in the space $s$ represent the states density already mentioned in this
section.  Now we see, that a given method leads to the determinant in
denomenator to set a density of paths contributing to the quantum
amplitude. In fact, this determinant is dictated by the uncertainty
principle  for the conjugate variables like coordinate and momentum.  The
merit of the average density valid for all paths, having the quantum
spreading fixed by the initial distribution, is evident at the caustic and
turning points.  Here, the zeroth derivative $R_{p}(t;s,0)=0$ provides for
the concentration of the classical paths, for which the Van Vleck's
determinats diverge $^{28,29,31}$.  However, our determinantal relation
still remains to be a good solution for the quantum amplitude.

\medskip

It is necessary to look more closely at the semiclassical recipe to gain
insight about a frequency modulation of wave-packet correlations.
The ${\cal S}$-correlator is written as a trace of the polarization
density matrix, see also eq  30,
$$
{\cal S}_{00}(t)=\mu^{2} \hbar^{-1}Tr\{\hat\rho(t)\}
\propto \int\limits_{s}{drdp}\;\rho(s,t).
$$
With initial condition $\hat\rho(0)=\hat {\rho}_{0}$,
a solution formally satisfies to
\begin{equation}
\hat {\rho}(t) =e^{-{i} t \hat H_{b}/\hbar}
\hat {\rho}_{0}  e^{{i} t\hat H_{x}/\hbar}
\end{equation}
The transport equation for the nondiagonal matrix element ${\rho}_{bx}(t)$
is immediately obtained with accuracy $\hbar^2$ to be sure containing the
description of quantum interference on the Franck-Condon transition:
\begin{equation}
\frac{\partial\rho(s,t)}{\partial t}+
\frac{p}{M}\frac{\partial\rho(s,t)}{\partial r}-
\frac{\partial U_{d}(r)}{\partial r}\frac{\partial\rho(s,t)}{\partial p}
-\frac{U_{d}(r)}{i\hbar}\rho(s,t)=0
\end{equation}
The difference potential $U_{d}(r)=U_{b}(r)-U_{x}(r)$ defines
corresponding resonant frequencies. An average potential is the half sum
of the molecular curves $U_{a}(r)=(U_{b}(r)+U_{x}(r))/2$.  The
equations of motion are
\begin{equation}
{\dot R} = {P}/{M} \quad \dot P = -{\partial U_{a}(R)}/{\partial R}
\end{equation}
So that, the  ${\cal S}$-correlations are represented as an integral of local
density with fast oscillating exponential prefactor in the space $s$
given by the expression,
\begin{equation}
Tr\{\hat\rho(t)\}=\int\limits_{s}{drdp}\;
e^{-{i}\int\limits_{0}^{t} d\,\tau U_{d}(R(\tau;s,t))/\hbar}
\rho_{0}(s(t))
\end{equation}
At the start, the function $\rho_{0}(s(0))=\rho_{0}(R(0;s,0),P(0;s,0))$
has a quite simple shape of Gaussian distribution. Its contour lines look
like the stretched ellipses. Then, the map exhibits a complicated topology
due to winding motion around a minimum of the average potential  and
its anharmonicity. Together with the fast oscillating functional for the
difference potential of the B-X states, it presents a challenge to
calculating the trace integral representation for 'topological
singularities'   of the turning point structure which is common for the
semiclassical approximations of wave mechanics $^{31}$.

\medskip

The solution to our problem is written down in Langrange coordinates
$S=(R,P)=(R(0;s,t), P(0;s,t))$. By virtue of Liouville theorem, the
phase volume in the jacobian transformation from Euler to Lagrange
variables is invariant, whilst molecular potentials are independent of
velocity. In the Langrange picture
\begin{equation}
Tr\{\hat\rho(t)\}=\int\limits_{S}{dR\;dP}
e^{-{i}\int\limits_{0}^{t} d\,\tau U_{d}(R(\tau;S,0))/\hbar}
\rho_{0}(S),
\end{equation}
the basic integrand becomes a function of R, P
$$
\hat\rho(t)=\exp(-i\Phi(S,t))\;\rho_{0}(S).
$$
To expand the path functional  $\Phi $  in the exponent to the second
order in departures from the point $q_0$,  we use a series

\begin{equation}
\Phi (s,t)=\Phi^0+\Phi^{0}_{s}\;(s-q_0)+\frac{1}{2}
\Phi^{0}_{s,s1}(s-q_0)(s1-q_0)+...,
\end{equation}
where
$$
\Phi^{0}=\Phi (s_0,t)=\int\limits_{0}^{t} d\,\tau
U_{d}(R(\tau;q_0,0))/\hbar
$$
and the path $s_0(t)=(R(0;q_0,t), P(0;q_0,t))$ obeys eq 45. By re-using
the vector and matrix notation again,  one can represent the final
expression   as a two-dimensional Gaussian integral,

\begin{equation}
{\cal  S}(t) =(\mu/\hbar)^{2} (\det {\hat \Pi})^{-\frac{1}{2}}
\exp \left (-i\Phi^0-\frac{1}{4}\;\vec U{\hat \Pi}^{-1}\vec U^{\dag}\right),
\end{equation}
where the vector $\vec U$  and matrix $\Pi$ are given by
\begin{equation}
{\vec U}=(\Phi^{\;0}_{r}, \Phi^{\;0}_{p});\;\; {\hat {\Pi}}=
\left( \begin{array}{cc}(\Delta R)^{-2} +i\;\Phi^{\;0}_{rr}/2
\;,&\quad\;i\;\Phi^{\;0}_{rp} \\
i\;\Phi^{\;0}_{rp}\;,&(\Delta R/\hbar)^{\;2}+i\;\Phi^{\;0}_{pp}/2
\end{array}\right)\;\;.
\end{equation}
The phase modulation is set by the function $\Phi^0(t)$. Additional
corrections in the exponent are responsible also for the decay  amplitude
giving generalized Debye-Waller factors for wave packet recoiling.
The phase and amplitude
formulae may be simplified  due to peculiar features of molecular curves,
short times  evolution etc. This approach would be of interest to address
Landau-Ziner transitions, and the  spectral lines structure from
impact center to static wings may be uniformly exploited.  An analytical
formulation of the globally optimal fields has some advantage of being used
to model the molecular wave packet correlations. On propagating in a space
the laser pulse might be tailormade to mimic the known optimal
fields. With this aim the high frequency filtering in optical guides,
nonlinear frequency dispersion, reflection from holographic film, gratings
etc. giving the right temporal and spectral trends  may be adapted to
perform the quantum control.

\medskip

As an example, we will discuss a chirping in  static wing of spectral
molecular line.  Accessed by the ultrashort pulse, the wave packet develops
in the average potential $U_{a}(R)$ between points connected by the direct
path. For a small $\tau$ (a femtosecond scale), these points are close
together on the trajectory
$$
R(\tau;R,P)=R+\tau P -(\tau^2/2M)\;(dU_{a}/dR)
$$
The difference potential in eq 46 can be expanded for the small
departures from the equilibrium point $q_0$, and we may rewrite the
frequency in $\Phi^{\;0}$ as a power series for a small $\tau$
$$
\begin{array}{c}
U_{d}(R(\tau;q_0,0)=U_{d}(R_0)-(\tau^2/2M)\;(dU_{a}/d{R_0})(dU_{d}/dR_0)=
\\ =U_{d}(R_0)-(\tau^2/4M)\;(dU_{b}/{dR_0})^2.
\end{array}
$$
Since the condition  $ dU_{x}(R)/dR=0$ (zero force)  should be met  in
$R=R_0$  (at a minimum of X curve), the result  is the quadratic chirp
lowering the pulse frequency with a correction coefficient of 0.25 to its
classical rate in  eq 34.  The correction of this sort is not
surprising for the classical - semiclassical - quantum correspondence.
Moreover, considering  more closely the phase and amplitude
modulation of wave packet correlations resulted from
$\vec U{\hat \Pi}^{-1}\vec U^{\dag}$   in eq 49,   one can find
additional corrections due to recoil from  the slope molecular curve. The
first non-vanishing terms of the power $\tau$ and $\tau^3$ in the exponent
are an exact match the quantum formulae  in the Condon model (see Sec. II
after eq 16).  It is worth stressing that the wave packet is found in
the coherent superposition of two resonance states, i.e. B and X, and
their mixing will reduce the chirp rate on the classically gained formula
in eq 34, because the molecular force acts only through the curve B (in
recoiling from the inner curve), and does not in the equilibrium point of
curve X.

\medskip

This argumentation must be revised further for returning trajectories on a
long time duration. In the case, our uniform treatment warrants a more
careful study that will be reported elsewhere. For completeness sake, the
basic ideas of quantum control should be examined in a quantum simulation,
which is favored to the semiclassical treatment,  if one cannot limit by
one classical path.  Thus, in the next section, we shall rely on the
quantum computation leaving aside a detailed numerical analysis of the
classical equations.

\section{ Numerical Simulations}

The quantum model eq 3 for the motion along one active I-I coordinate
and electronic transition belongs to the polynomial class of numerical
complexity of  $N$ linear equations, where N is the number of  grid
points.  There is more that one recipe with which to perform their computations
$^{32}$. The QR diagonalization requires the $O(N^3)$ operations.  In
specific cases, the QR algorithm, as a stable spectral decomposition, can
be used for the wave packet propagation.  The time implicit symmetrical
schemes of Gaussian elimination $^{33}$ necessitate $O(N)$ operations. The
symmetrized split operator method  does $O(Nlog(N))$ steps based on the
fast Fourier transform (FFT) $^{34}$.  The FFT propagates the wave
packet similar to Feynman's path integral and alternates the coordinate and
momentum representations involving the thorough dynamical picture in hand.
These methods are in good agreement.  The practical value of FFTs or
Gaussian elimination is that they enable to get rid of the matrix
diagonalization.

\medskip

The absorbing boundary condition has been used in the numerical
simulations. The imaginary negative optical potential discriminates the
outgoing wave packet at outer edge of the numerical grid and has no effect
on the bound states. This feature is visible in  Figure 2, where the molecular
dynamics following impact resonance interaction on B-X transition is
shown.  The ground state  replica transforms into  a wave-packet
which is scattering above the molecular
dissociation limit and binding below it.  The scattering states, shown in
the foreground of  Figure 2, are involved due to impact photodissociation
(i.e., the $\delta$-like pulse action).  This branch is not mixed with the
bound-bound transitions responsible for coherent recurrences, while the
wave packet recoils against the molecular walls and  oscillates
spreading  for the B state anharmonicity.

\medskip

The resonance scattering  radiation  in the wake of ultrashort pulse
proceeds in two stages:  a fast monotonic decay due to direct
photodissociation gives way to a "non-regular" reviving.  This picture is
shown in  Figure 3 for the fundamental overtone of Raman intensity. Since
a damping out is assumed to be small, the wave packet correlations hold
during many molecular periods.  The amplitude and phase  profile will
follow in reverse order.  Thus,  Figure 3 showing  its fundamental overtone
intensity  must be read " from right to left".  The phase conjugate
scattering signal in  Figure 4(a) has been designed to drive the spread waves
towards $t_d$.  The target time $t_d$ exactly represents the starting
point of wave-packet correlations.  Furthermore,  on going back in time their
envelope ${\cal S}(t_d-\tau)$ is not delta correlating as a white noise
signal.

\medskip

There are two regular trends: the  negative frequency chirp, which
has been explained with the semiclassical arguments, and amplitude
growing  to the pulse end.  The last feature stemmed  from  the
reversibility theorem can be also understood qualitatively.
Since the wave packet  is spreading between turning points, the
result is a small amplitude almost everywhere in the classically
accessible Franck-Condon region.  The optimal field on its start is set by
a projection of the spread wave packet to the X state.  The S-correlations
undergo fast oscillations due to multiple recoils of the wave packet
against molecular walls and its spreading accompanied by decreasing
amplitude.  Evolving back in time, the wave packet becomes more regular,
and its shape is recovered and copies well localized replica of the ground X
state.  Focussing the wave packet in the vicinity of the inner turning
point results in a further rise of its amplitude \footnote{A clear analogy
is given by a point of intense radiation  behind burning-glass exposed to
sun.}. The projection to the ground X state grows together.  This explains
why the globally  optimal field behaves  in such a way in  Figure  4(a).
Induced resonance transient given by eq 4 is shown in  Figure   4(b). The 1
ps square gate of optimal field  creates the transient spike at delay time
1.5 ps.   The temporal full width at half maximum (FWHM) equals $T_{w}=25$
fs that is in good agreement with the kinematical overlap lifetime
$\tau_{rec}=15 fs$.

\medskip

This semiclassical estimate is astonishingly  accurate although less then
quantum computed. There are two main reasons of this distinction.  The
wave packet interference is absent in the classical kinematics and the
dispersion of excited state is not taken into account.  Thus, the
ignorance of quantum wave nature underestimates the width of the coherent
peak.  To cure the situation, the semiclassical uniform approach, which
copes with initial quantum distribution, classical motions, caustic
and turning points, may be applied.

\medskip

For definiteness sake, setting $T_g=50$ fs square gate of optical target $G$
at the delay time 1.5 ps, we now can consider a realistic optical transient.
Averaging the globally optimal envelope over $G$ washes out the high
frequency components as shown in  Figure 5.  However, the average pulse shape
is yet endowed with a chirp.  This feature is exhibited in  Figure  6 (a), where
Wigner spectrogram is shown. The Wigner function of the optimal fields is
defined as
$$ W(t,\nu)= \int
\frac{d\tau}{2\pi\hbar} e^{{i}\nu \tau/\hbar}
{E}(t+\frac{\tau}{2})\stackrel{*}{E}(t-\frac{\tau}{2}).
$$
Picosecond square gate envelope and periodic boundary condition have been
used in evaluating the spectrograms.  The frequency-time maps manifest a
striking correlation between the optimal fields and  wave packet
evolution.  For instance, the envelop of allocated duration $T_{p}=1 ps$
consists of more shorten  subpulses.  Their number is associated with the
number of vibrational cycles between turning points during $T_{p}$.  For
every recoiling from inner molecular core the wave packet gives rise to a
subpulse.  The contour levels of the two-dimensional Wigner spectrogram are
taken at one-half a maximum height.  A slope and concavity characterizes
respectively a linear and quadratic frequency chirp of subpulse.  The
semiclassical reasoning relate the chirping to wave packet velocity and
acceleration.  The chirp is in fact linear because time duration of pulse
$T_p$ (in picosecond range) is much more than the recoil lifetime $\tau_{rec}=15$fs.
Thus, the wave-packet is far from turning points and does not accelerate
most of the time.   The spectral FWHM in  Figure 6 (a) is
about $\Delta\nu\approx500 \; cm^{-1}$ being consistent with the
reciprocity relation $ \Delta\nu=  2\pi\hbar/\Delta T_g $ $^{35}$ .

\medskip

A minor decoherence of the objective as shown in  Figure  6b is obtained in
our numerical simulation.  The temporal FWHM is about 60 fs being
10 fs over the square gate target width $T_g=50$ fs. The distinction
stems from the wave packets quantum dispersion.  The wave packet
dynamics is displayed in  Figure   7 and  Figure  8.  The resonance frequency is
$19050 \; cm^{-1}$ above the ground X term; the excitation bandwidth is
about  $500 \; cm^{-1}$ (FWHM). Thus, the wave packet B is created below
the molecular continuum  to spread  from $2.2\AA$ to $4 \AA$ not far from
its dissociation limit. Given intensity of transitions, the depopulation
of state X is involved to the pulse end, although  the X state
localization is not affected at the saturation.  The optical transient
rises and falls, following the wave packet recoils from the inner core.
The spatial variance of a maximum focused wave packet  at target time
reaches 0.04 $\AA$, being slightly less than the original localization in
agreement with our qualitative reasoning.  The result is robust to slow
sinusoidal or Gaussian aberrations of optimal fields and akin to known
stability of phase conjugate fields due to the wave speckle
structure $^{36}$.

\section{Conclusion and Future Perspectives}

The numerous possible applications in femtosecond spectroscopy are rested
on the concept of coherence. The proceedings of Femtochemistry III
Conference in Lund testify (see the special issue of the Journal of
Physical Chemistry), that this realm of knowledge is flourishing and
another lines of  researches have now been opened.  As an example
we refer the field of deemed quantum
computers$^{37-39}$. Here, controllable coherence of logic gates could
boost performance of massive parallel computation exponentially with
number of quantum nodes.  These nodes may be constituted from clusters or
single molecules and their resonance optical transitions must act in
concert with vibrational wave packets captured by "molecular cavities". A
run stage of a quantum computer, its coherence may not be affected by
environment $^{40}$. Thus, the danger of dephasing requires the optical
coherence to be actively controlled according to a state-mapping protocol.
From this point of view, the reversibility theorem stated in our paper
exemplifies a paradigm of controllability expected for wave packets
dynamics and optical transients.  In fact, the fundamental connection
between physical laws and computations based on time reversable
Hamiltonians has been addressed yet by R.Feynman $^{41}$. The
computational quantum  networks  requires the reversable transmission
among distant nodes $^{42}$.

\medskip

The central aim of this paper is to set a globally optimal solution to a
maximum squeezing  the optical coherence of a single molecule.  The
sufficient conditions of recovering coherence have been stated in the
reversibility theorem.  Our time reversed fields have been tested in
the model computation of $I_2$ B-X transition confirming effeciency of
the phase conjugate resonant Raman scattering for quantum control.  The
result is not limited only to diatomic molecules in gas or condensed
states, but is valid for the quantum control of electron wave packets as
well $^{43,44}$.

\medskip

The four-wave mixing interaction and related stimulated Raman emission is
deserving of mention. These optical transients would make it possible to
automate the quantum control to be held both for bound states and
molecular continuum.  To take one example, ultrashort transform limited
pulses driving bound-free transitions result in the photon echo of
photodissociation $^{45}$, to mean  an automated controlling wave packet
motion.  Photon echoes $^{19}$ imply that dynamical holograms are written
on resonance Franck-Condon transitions.  But it takes more strong laser
fields  to induce at least the third order nonlinear polarization. A
self-adaptive "quantum mirror" has been devised $^{36,46}$ to reverse wave
evolution on curing the optical aberrations of a noise wave front
propagated trough inhomogenious media. Moreover we have seen that the
phase cojugate (time-reverse) Raman scattering negates also delocalization
of molecular wave packets.

\medskip
The use of the phase conjugate signals presents an intriguing illustration
of how the original quantum state is restored in the matter-wave
interferometry. An uniform semiclassical formula has been derived for the
optimal field envelop. The simple estimates are contrasted with quantum
simulation. We conclude that femtosecond  coherence of a single molecule
must exist in varied conditions (even on resonant transitions saturated to
the pulse end).  Robustness of our molecular magnetron design has been
demonstrated also.  We propose to test experimentally the molecular
magnetron design verifying the  quantum holography  directly on observing
delayed sparks of femtosecond fluorescence which is excited by a more long
picosecond  pulse in optically thin sample.

\section{Acknowledgments}
We thank K.R.Wilson for reprints on controlling the future of matter
available before their publication. The work benefited from useful
discussions with S.R.Hartmann, J.Jortner, M.Chergui and I.Averbukh.
Sincere thanks to the referees  of our manuscript for their critical
reading, advices and editorial help in getting the paper to readable form.
One of the authors (A.Rudavets) is especially indebted to V.Aquilanti,
B.Soep, J.Chergui, K.Kulander, V.Sundstroem for a kind hospitality
extended to him to visit quantum control meetings around the world.
It was supported by the Russian Foundation for Basic Researches grant
96-02-18760.

\begin{figure}
{\includegraphics[width=6 in]{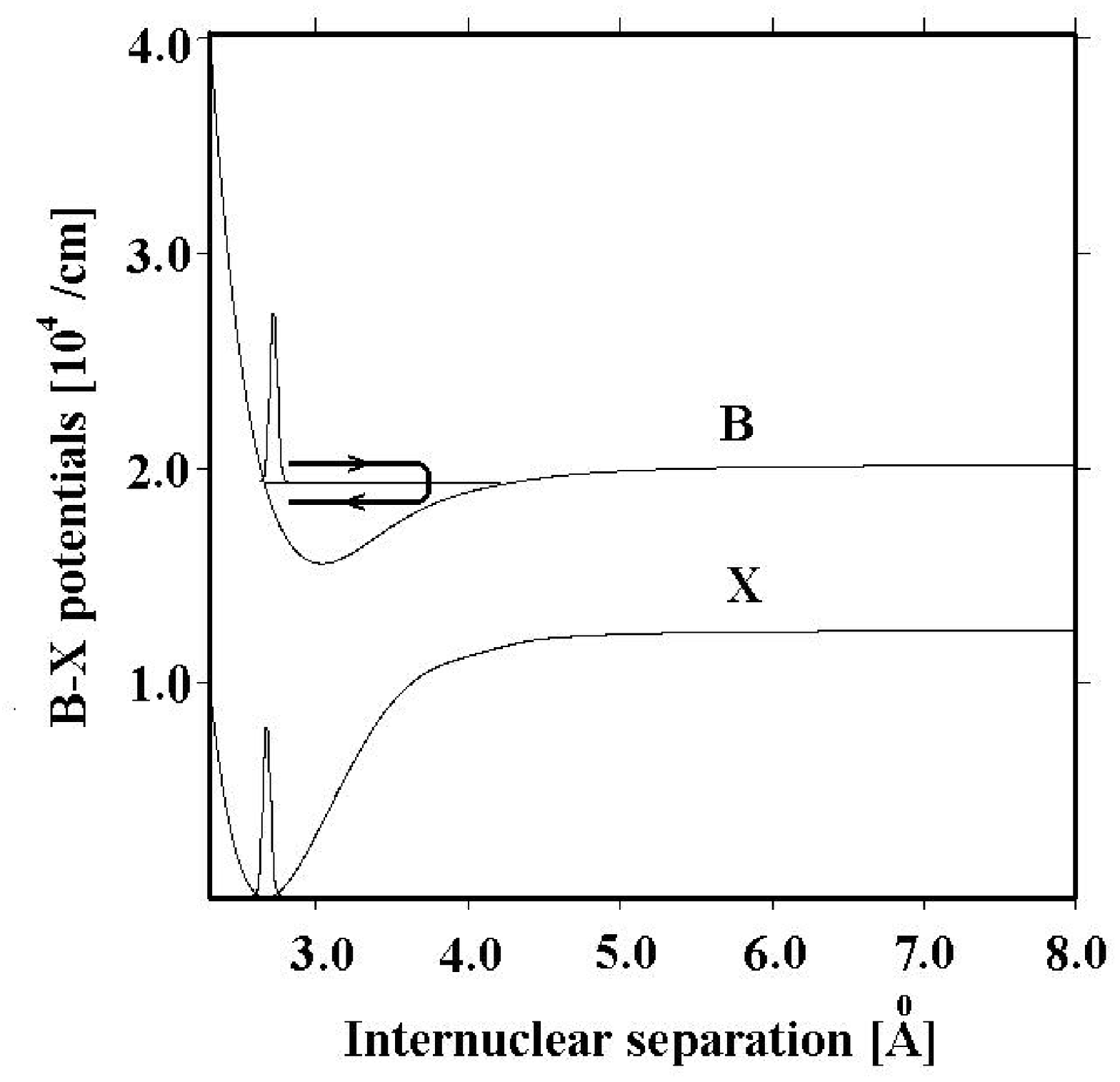}} \caption{The resonance curves of
$I_2$ molecule [20]. Our design represents the iodine magnetron.  A tailored
laser pulse, as shown by the point-filled box, excites the vibrational wave
packet.  Then, the wave packet is reflected from outer curve and moves back
to focus on the inner Franck-Condon region.  The result is a tight overlap
with the ground state X.  The maximal optical polarization must be followed
by the resonance fluorescence burst  at target time.}
\end{figure}

\begin{figure}
{\includegraphics[width=6 in]{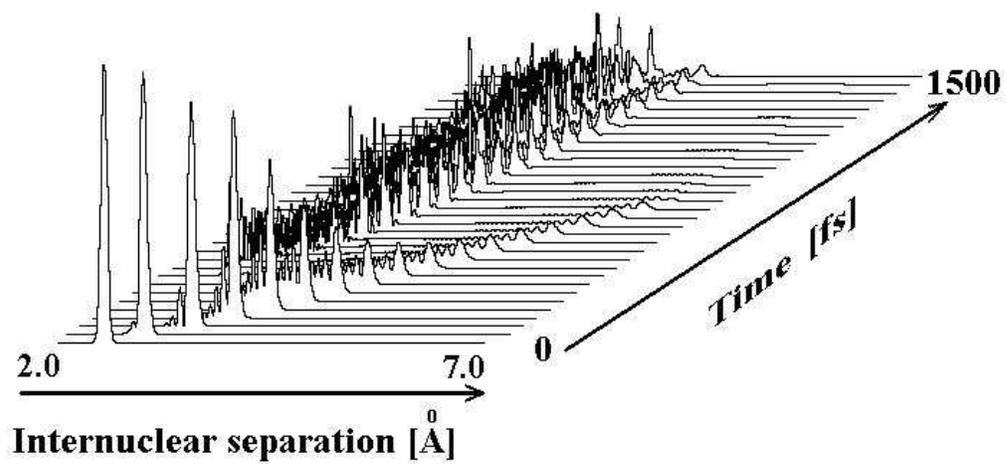}} \caption{The picture shows the
nodal structure of bound wave packet ranging between $ 2.2 - 4 \AA \AA $.
The wave packet fringes are due to interference of the wave packet in virtue
of multiple collisions against the molecular curves.}
\end{figure}

\begin{figure}
{\includegraphics[width=6 in]{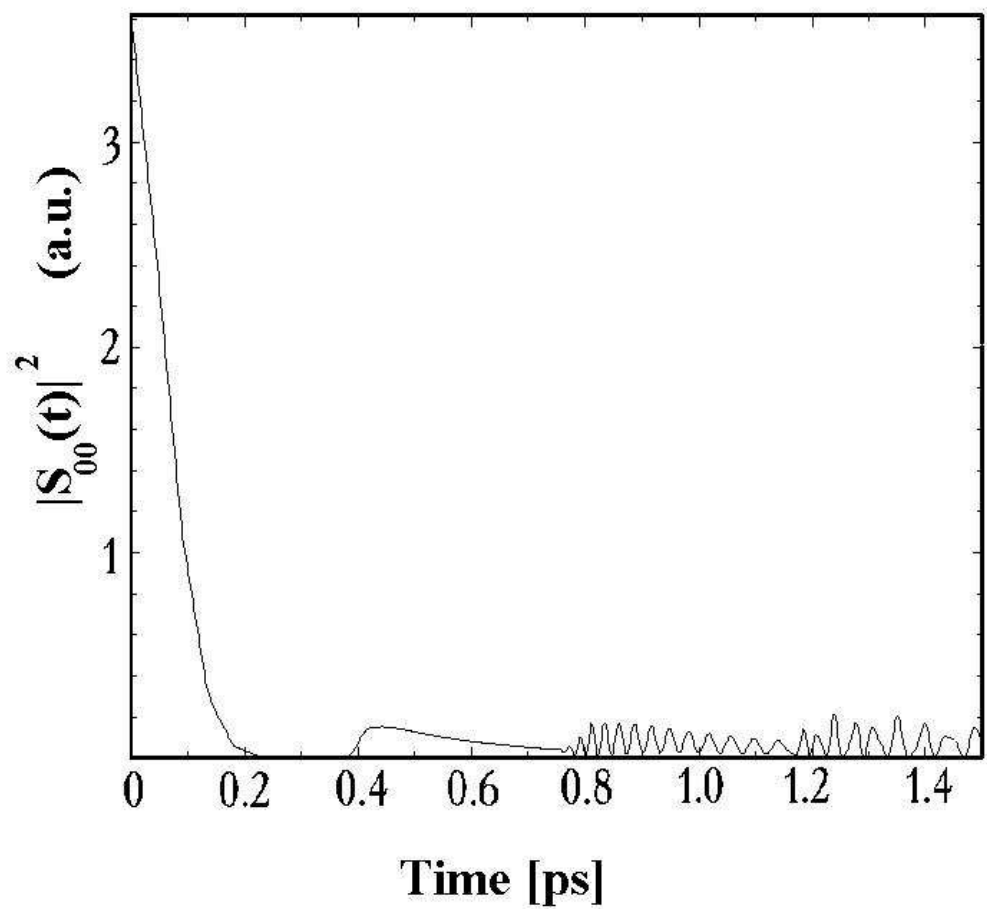}} \caption{Intensity  of fundamental
overtone of resonant scattering as a function of time. This resonance
transition is induced by the photon impact (the delta-like pulse).  The
scattering B states contribute to monotonic free decay signal first few fs.
The bound states interference, present in the background in  Figure   2, is
responsible for reviving the resonance Raman scattering after 320 fs delay,
which is matched with a molecular period (at 570 nm in B state).}
\end{figure}

\begin{figure}
{\includegraphics[width=4 in]{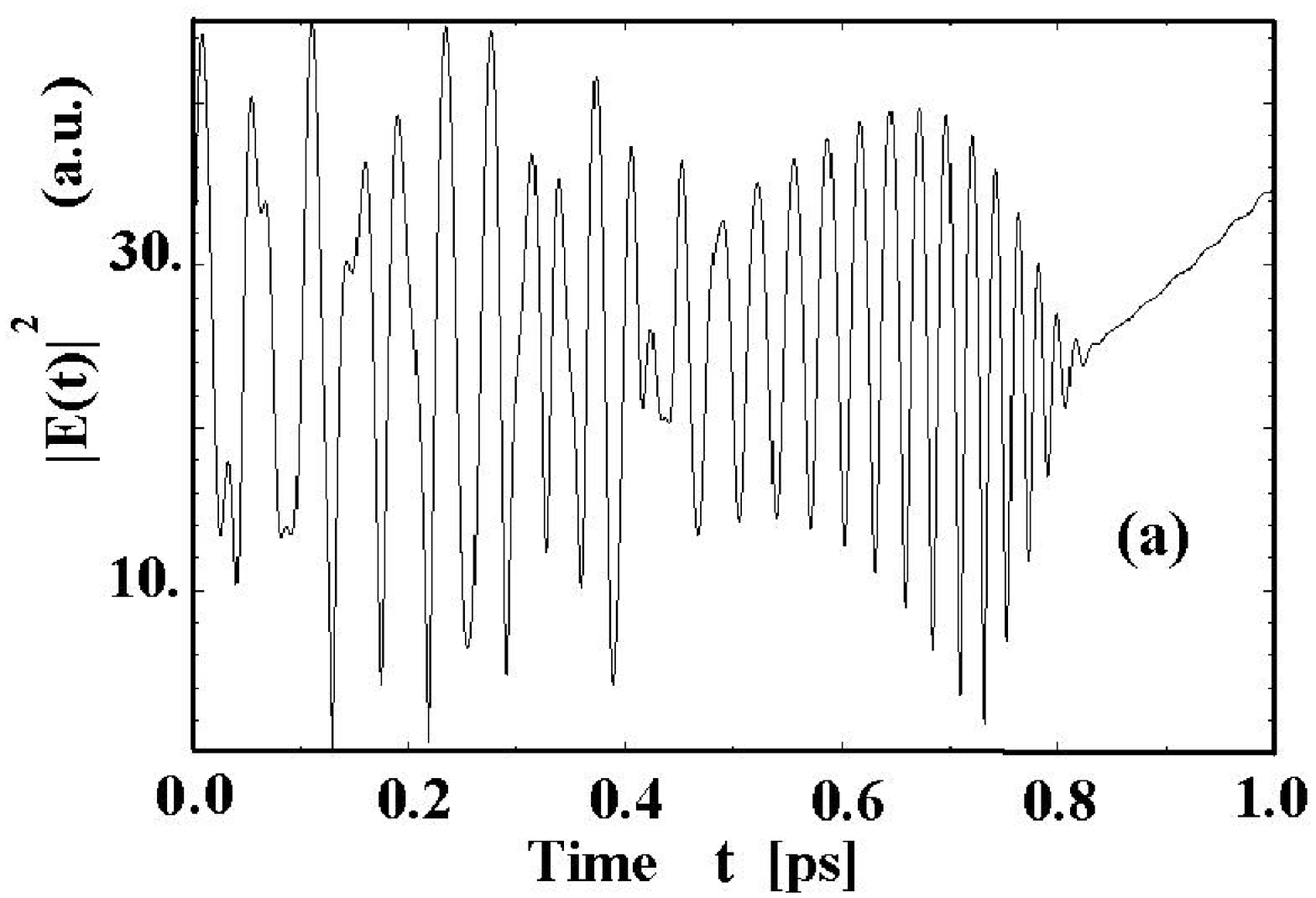}} {\includegraphics[width=5
in]{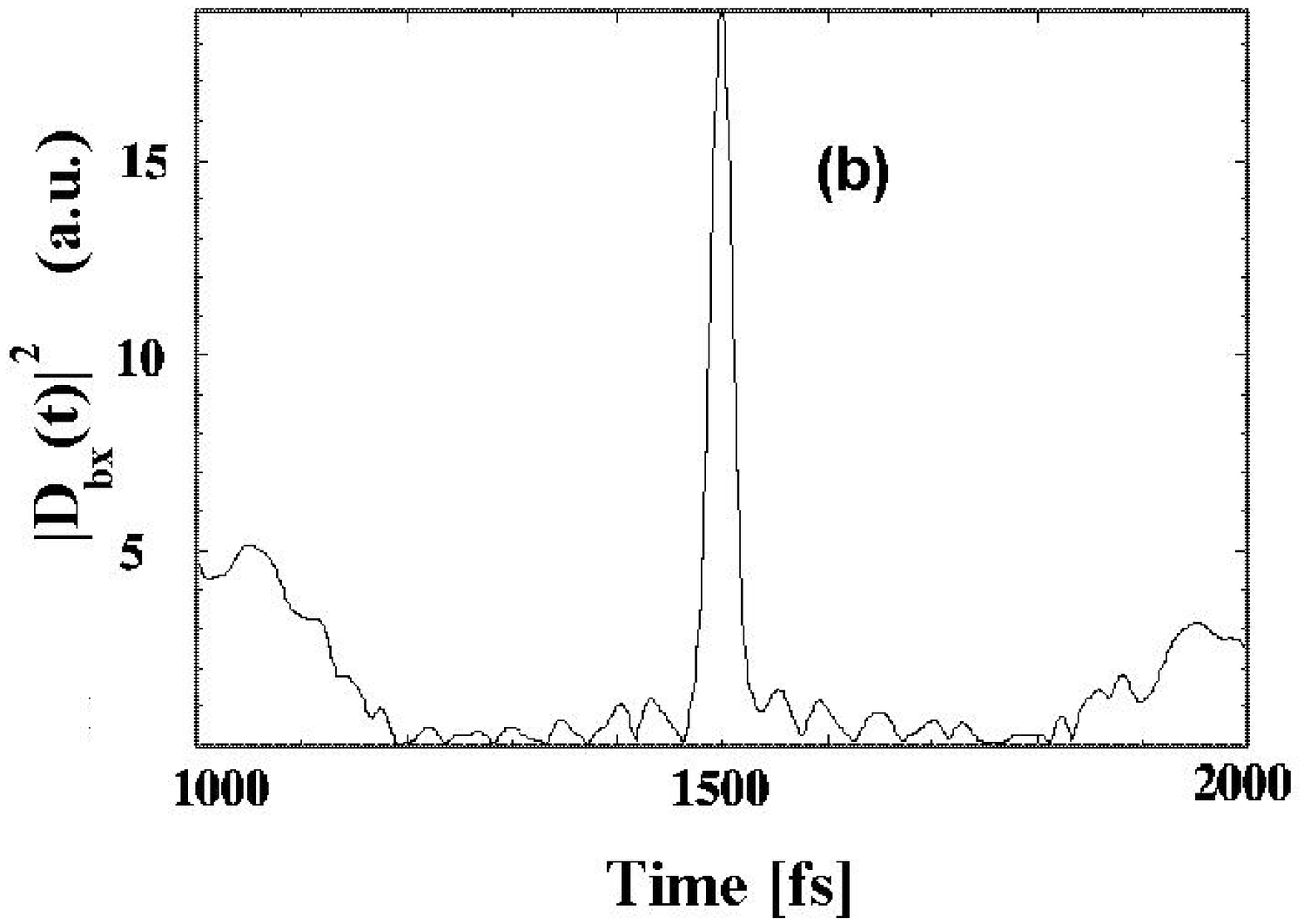}} \caption{The square 1 ps gate of the optimal field in (a).
This optimal field creates the dipole transition moment shown in (b). The
accomplishment of our coherent objective is evident at target time equal to
1.5 ps.}\end{figure}

\begin{figure}
{\includegraphics[width=4 in]{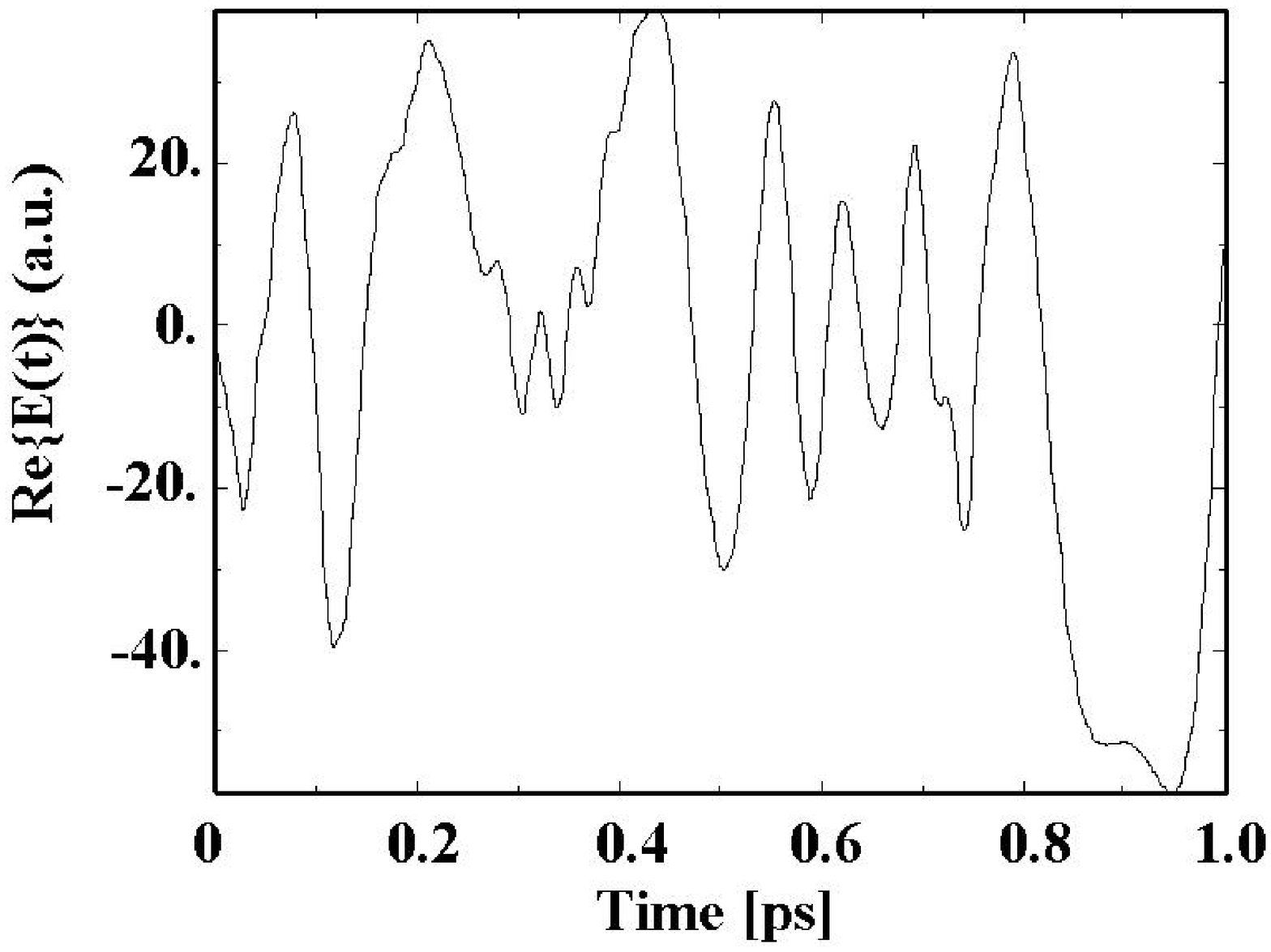}} {\includegraphics[width=4
in]{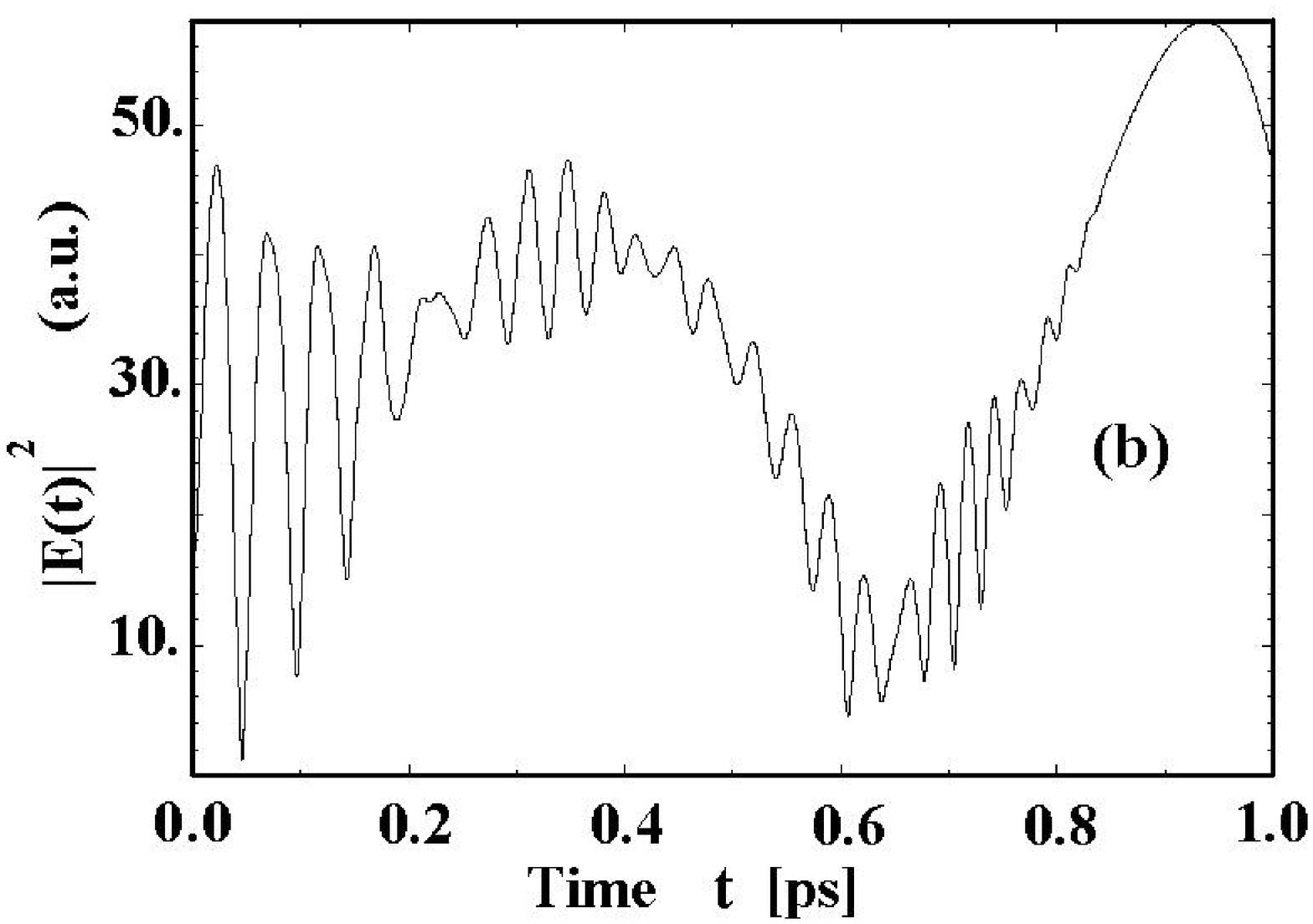}} \caption{The intensity of optimal field in (a) and its real
part in (b). This optical pulse is designed to create 50 fs square gate
optical coherence at target time 1.5 ps.}
\end{figure}

\begin{figure}
{\includegraphics[width=4 in]{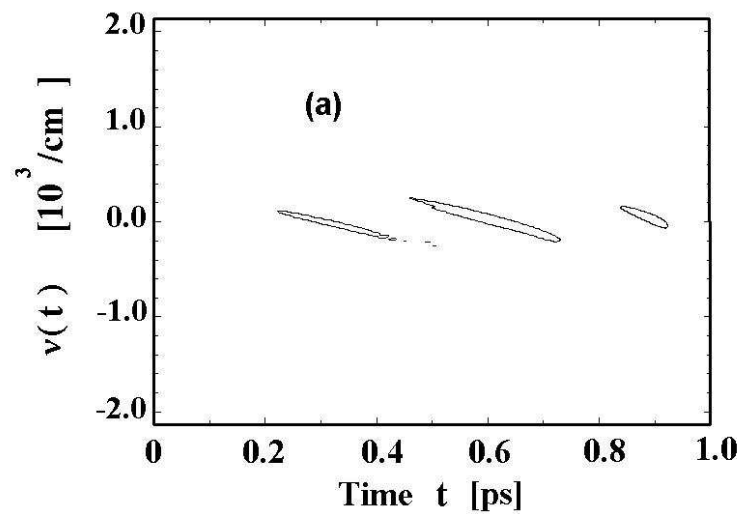}} {\includegraphics[width=4
in]{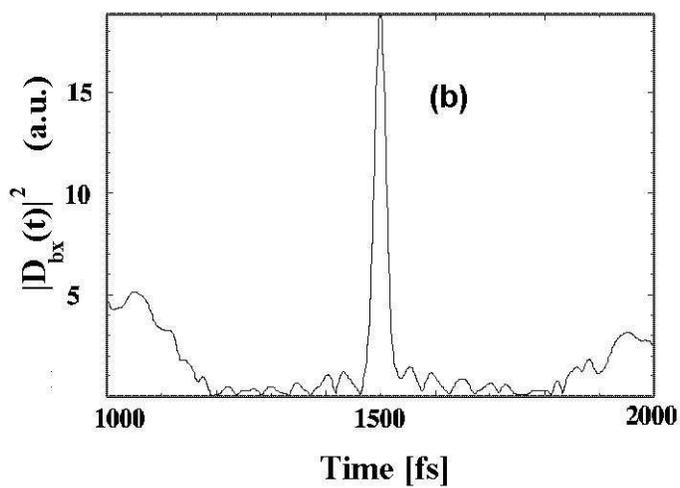}} \caption{The frequency-time plot of the globally optimal
pulse for the 50 fs square gate target is given by the contour map at half a
maximum of Wigner spectrogram in (a). The coherent transient in (b) is
created by the optimal field.  The temporal FWHM equal to 60 fs fits well
the 50 fs square gate target.}
\end{figure}

\begin{figure}
{\includegraphics[width=4.5 in]{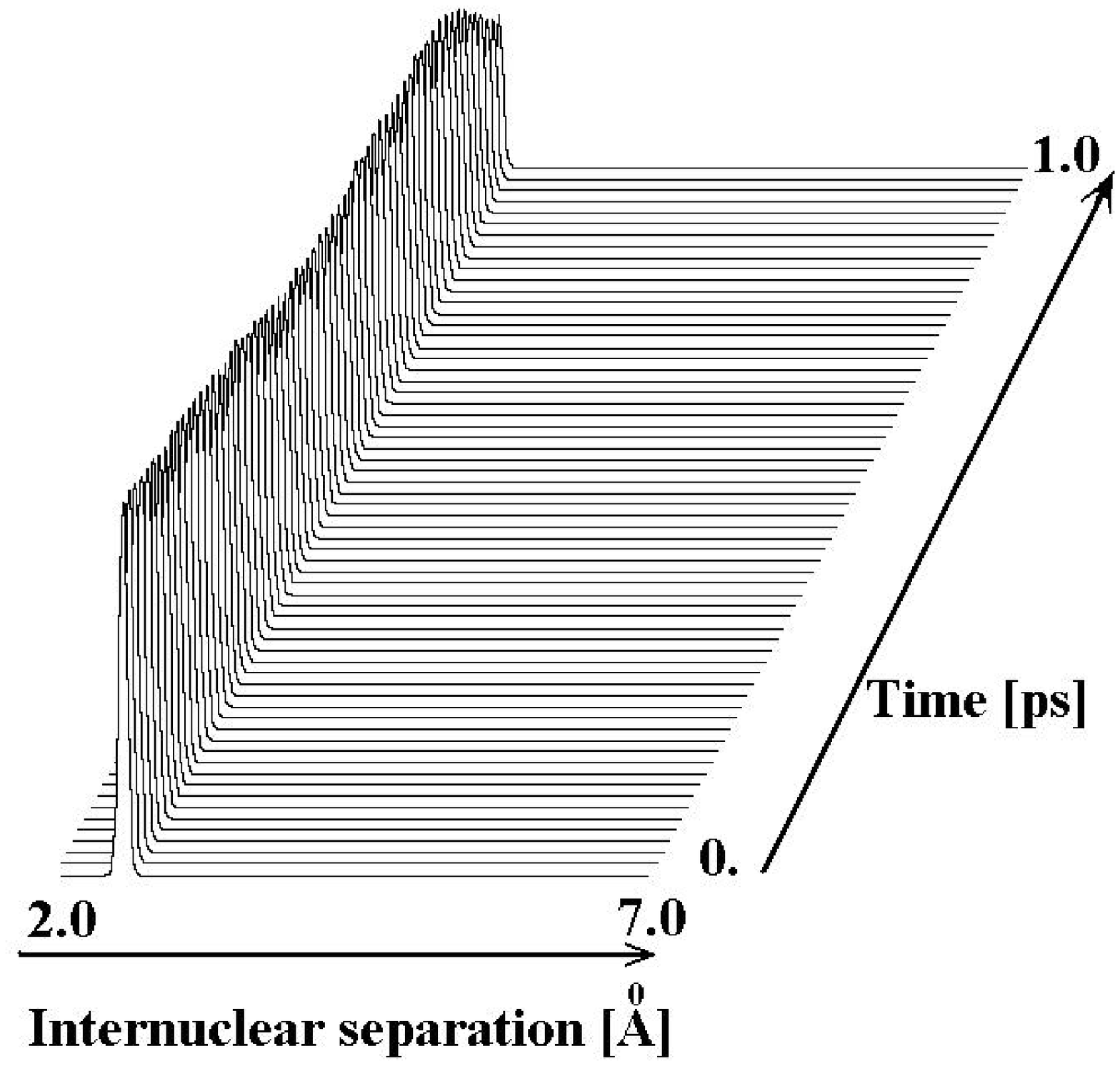}} {\includegraphics[width=4.5
in]{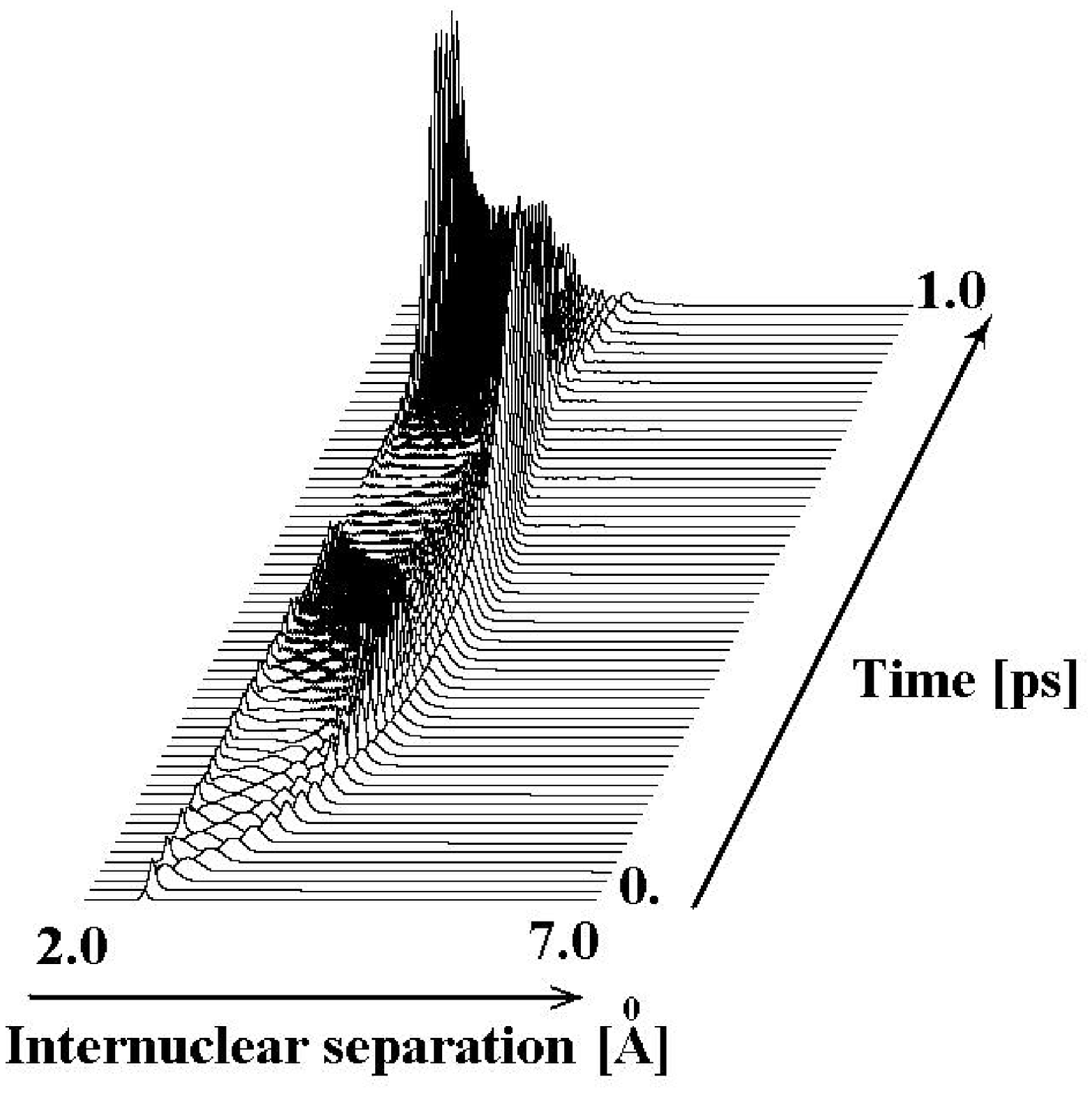}}\caption{The tailored field shown in  Figure 5 excites the
delocalized vibrational wave packet in (a). The X state in (b) holds its
initial spike shape centered at 2.66 $\AA$  with variance 0.05 $\AA$.}
\end{figure}

\begin{figure}
{\includegraphics[width=6 in]{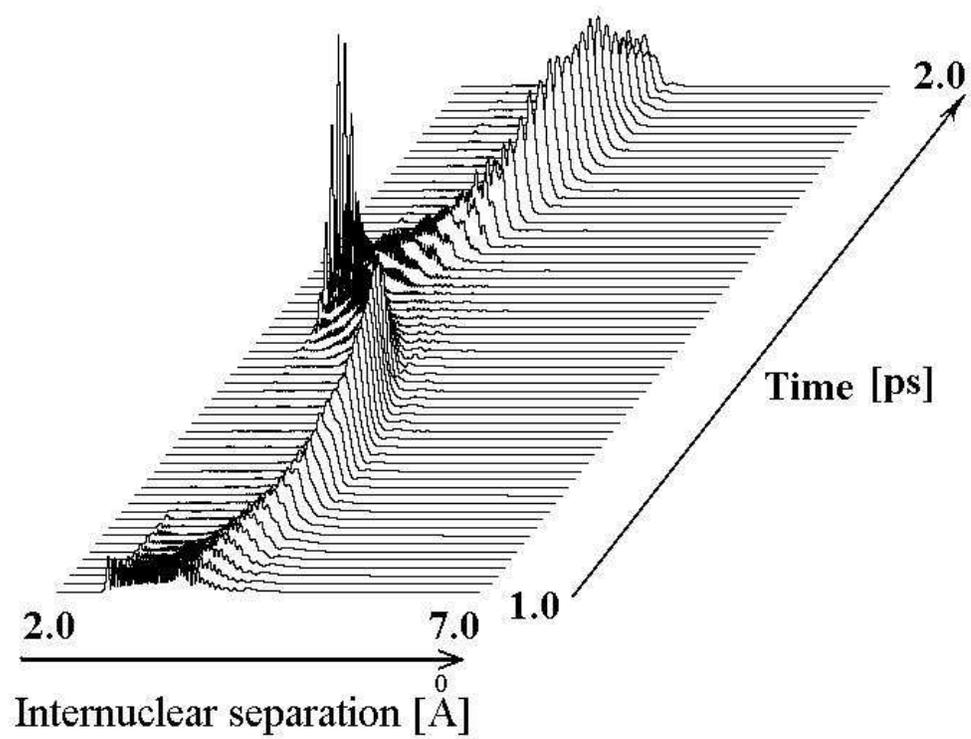}} \caption{ The wave packet B free
propagates after the pulse to a tightly squeezed state at the inner turning
point to overlap well the ground state X at target time 1.5 ps.}
\end{figure}

\end{document}